%% file: main.tex
\documentclass[runningheads]{llncs}
\usepackage{graphicx}
\usepackage{tabularx}
\begin{document}
\title{Best practices\\for software maturity improvement:\\a G\'{E}ANT case study \thanks{This work is part of a project that has received funding from the European Union's Horizon 2020 research and innovation programme under Grant Agreement No. 856726 (GN4-3). \protect\\The scientific/academic work is financed from financial resources for science in the
years 2019-2022 granted for the realisation of the international project co-financed
by the Polish Ministry of Science and Higher Education.}}
\titlerunning{Best practices for software maturity improvement: a G\'{E}ANT case study}
%
\author{Bartosz Walter\inst{1} 
\and
Branko Marović\inst{2} 
\and
Ivan Garnizov\inst{3}
\and
Marcin Wolski\inst{1}
\and
Andrijana Todosijevic\inst{4}
}
\authorrunning{B. Walter et al.}
%
\institute{PSNC, Poznań, Poland
\email{\{bartek.walter,marcin.wolski\}@man.poznan.pl}
\and
University of Belgrade, Belgrade, Serbia,
\email{branko.marovic@rcub.bg.ac.rs}
\and
Friedrich-Alexander-University of Erlangen-Nürnberg, Erlangen, Germany,
\email{ivan.garnizov@fau.de}
\and
AMRES, Belgrade, Serbia,
\email{andrijana.todosijevic@amres.ac.rs}
}
\maketitle              
\begin{abstract}

Maturity models for software indicate the key areas that contribute to quality improvements. They usually combine technical, organisational and human aspects relevant for effective software development, to focus the efforts and draw the direction for optimisations.
In this paper, we present the process of defining best practices that support the G\'{E}ANT Software Maturity Model (GSMM), aligned to the needs of a distributed, innovation-driven, pan-European organisation. Based on the identification of specific goals relevant for G\'{E}ANT and a preliminary maturity assessment, we created a catalogue of best practices that help the software teams to attain the goals defined in the GSMM.

\keywords{maturity evaluation \and best practices \and software process improvement \and SPI}
\end{abstract}
\input{intro}

\input{relwork}

\input{geant}
\input{process}
\input{bestpractices}

\input{results}
\input{conclusions}

\bibliographystyle{splncs04}
\bibliography{gsmm}

\end{document}

%% file: intro.tex
\section{Introduction}
\label{sec:intro}

Managing software process improvement endeavours is a complex challenge that usually involves the effort of several teams and individuals. In particular, it is the case for large organisations focused on innovation, with an established culture of diversity and openness~\cite{Stanisavljevic2018}. The process improvement usually requires identification of factors relevant in a given context, setting attainable objectives, defining metrics for tracking the progress, but also coordinating the efforts in various areas: technical, human and organisational.

The concept of maturity, which captures the capability of an organisation to deliver high-quality products, is widely accepted as an effective method of improving the processes. Maturity models provide frameworks capturing the essential dimensions of quality that are relevant in a given context, to set objectives and propose methods of addressing them. Several models have been proposed for software development, both generic~\cite{Paulk1993,Burnstein1996-tmm} and tailored~\cite{Fontana2015}. 

However, defining the model is only one side of the coin. Apart from identifying the goals and defining the metrics, the subject teams and individuals also need guidance, support and actionable recommendations on how to work to attain the objectives. Although it is quite feasible to directly implement the model in uniform and hierarchical organisations, for internally diversified and independent structures it could be a difficult task.

This also is the case of G\'{E}ANT, a pan-European organisation, established and funded under several EU programmes for the development and operation of a fast, reliable networking environment for research and education, which offers software-based services to various end-users, including students and researchers. It involves many independent software teams that are free to define their processes and obliged only to adhere to the common organisation-wide recommendations.

In this paper, we report on how a custom maturity model for G\'{E}ANT could be supported by a catalogue of best practices. The catalogue guides on how the specific objectives of the model could be addressed and implemented by the G\'{E}ANT software teams.

The paper consists of seven sections. In Sec.~\ref{sec:relwork} we report a literature overview; in~\ref{sec:geant} we shortly introduce the G\'{E}ANT organisation and its specifics concerning software development. In Sec.~\ref{sec:process} we present the entire process of maturity improvement, from defining the model to constructing the catalogue of best practices. Next, in Sec.~\ref{sec:practices} we present how the best practices are described, formatted and presented in a catalogue.
Sec.~\ref{sec:results} reports the early results of the evaluation, and the Sec.~\ref{sec:conc} provides concluding remarks and the summary.


    

%% file: relwork.tex
\section{Related work}
\label{sec:relwork}

Maturity of software organisations is a topic widely explored in literature. Several software-related maturity models have been developed that refer to specific areas and scopes, e.g. software process capability models (CMMI)~\cite{Paulk1993}, software analysis~\cite{covey2005creation}, operational management~\cite{renken2004developing} or business process management  ~\cite{Bruin2007}. Although maturity is usually related to traditional methods of software development, the concept of maturity has been also tailored for agile approaches: Agile  Maturity  Model  (AMM)~\cite{Patel2009} defines levels  of agility, which address the common agile practices and  values starting from basic ones, e.g., planning and requirements management, up to managing uncertainty and defect prevention.

A number of maturity models have been also proposed in EU-funded projects. They address various areas that could be partially relevant in a software-related context, e.g., communication~\cite{Muszynska2018} or selected education techniques\footnote{\url{https://embed.eadtu.eu/}}. However, they usually focus on a single, selected dimension of the project.

As a consequence, although numerous models exist, they still need to be merged, customized or redefined to reflect the specific requirements and settings and to embrace all areas relevant for software development. To respond to this, in previous papers~\cite{Stanisavljevic2018} we presented a preliminary version of the GSMM, a maturity model dedicated for the G\'{E}ANT organisation, along with recommendations on how to define models and implement them~\cite{eurospi2019}. 


 
Effective implementation of the objectives and goals defined in maturity models requires also adequate guidance and recommendations. They could take the form of best practices that are well-founded on both the experience and the existing knowledge, and are applicable in the relevant context. This approach is widely adopted in software engineering, e.g., in SWEBOK~\cite{swebok}.
Catalogues  of  such  practices dedicated to specific areas have been  proposed  by  various  authors.  Gamma et al.~\cite{gamma1994design} expressed the collected experience in designing object-oriented software as a set of design patterns. They documented key design templates  and  presented them in the form of a catalogue of abstract structured recommendations. Also, anti-patterns have been defined, capturing practices that should be avoided~\cite{brown1998antipatterns}. Similar efforts have been undertaken also in several other software-related areas, e.g., testing, documentation etc.

Ambler~\cite{Ambler}, based on his observations, emphasized the importance of the context in the analysis of best practices. He argued that most practices are not applicable in all cases, and they needed to be either adapted before being applied or to be implemented only in specific environments.

We believe there is still a need for presenting custom, organisation-specific models for improving software maturity, supported by structured, practice-originated experience.

%% file: geant.tex
\section{Background}
\label{sec:geant}


G\'{E}ANT is a pan-European project focused on the development and maintenance of e-infrastructure and services for the research and education community. It operates the backbone network and associated services interconnecting national research and education networks (NRENs) across Europe and enables their collaboration. Also, it is a distributed, innovation-oriented organisation involving participants from many countries and organisations that develop and maintain network-based products and services, frequently based on dedicated or customized software. G\'{E}ANT portfolio comprises currently 30 software projects: some are used directly by G\'{E}ANT; some more are shared or used by NRENs; yet, others contribute to wider open-source communities. 

Members of the software teams have  specific working arrangements: they simultaneously work for G\'{E}ANT and their native organisations, can be simultaneously involved in several projects, are geographically distributed, and are placed in different cultural and professional backgrounds. The teams share the common software development framework provided by G\'{E}ANT, but are allowed to choose and customize their processes, methodologies and approaches. Their developments are focused on innovative and often prototypical applications for the high-performance network, based on the novel and often federated services.

A previous analysis of the G\'{E}ANT software development practices~\cite{Wolski2017} showed that there is a need within the G\'{E}ANT software development community for optimisation of the software development processes. This need could be addressed by providing the software teams with guidance on adopting and using software development methodologies and practices effectively and efficiently.

The motivation for the establishment of a software maturity model for G\'{E}ANT was to determine properly the improvements schema for software teams. Moreover, the practical appliance of maturity model would align the improvement effort with governance frameworks, commonly approved models, and with industry practices. The apparent attractiveness and popularity of the maturity model in the management of the software process and, more broadly, IT management, has contributed to our effort to develop a maturity model specifically for software process improvement (SPI) within G\'{E}ANT, while respecting the seven suggested requirements for the development of maturity models~\cite{Becker2009}.

The G\'{E}ANT Software Maturity Model (GSMM) has been designed to achieve two primary goals: (1) to capture key practices that already help the teams to successfully deliver software, and (2) to identify areas for further improvements that could be applied by the teams~\cite{eurospi2019,Stanisavljevic2018}.

The extracted practices can then be shared, adapted and applied by the teams to streamline and align software development processes and governance. The leading factors considered include the lasting nature of G\'{E}ANT, its products and services, distributed nature of conducted collaborations and the existence of many practices that have been established for some time. 

The resulting model consists of categorised software engineering topics and processes into five key thematic areas, referred to as target areas (TA), namely: requirements engineering; design and implementation; software maintenance; quality assurance; and team organisation. These target areas were elaborated by providing specific content to the maturity model, with each one consisting of several specific goals (SGs) that capture sub-objectives and related activities.

%% file: process.tex
\section{Process of defining the best practices}
\label{sec:process}

    
Best practices are commonly accepted procedures that aim at accomplishing certain objectives. They are applicable in a given context. They are considered as the gold standard for attaining specific objectives.
 
\begin{figure}[h]
 \centering
 \includegraphics[scale=0.5]{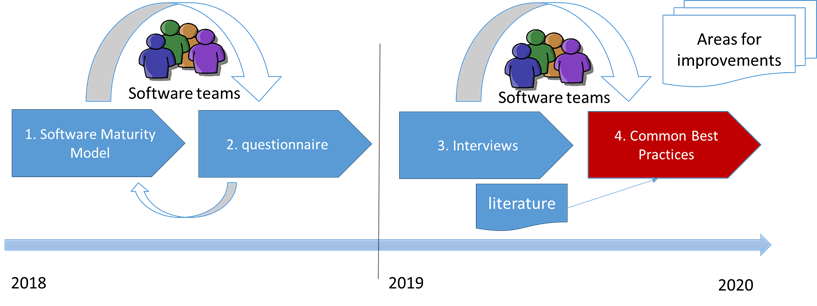}
 \caption{The process of defining best practices adopted for GSMM}
 \label{fig:process}
\end{figure}

The process of defining the best practices comprised the four main steps (see Fig.~\ref{fig:process}):
\begin{enumerate}
\item Defining the software maturity model which would determine our improvements schema;
\item Preparing the questionnaire which would be consistent with the structure of the software maturity model;
\item Interviewing selected G\'{E}ANT software teams to collect both qualitative and quantitative information and opinions on how the specific goals are addressed;
\item Determining the set of Common Best Practices (CBPs), based on the results of the survey, literature review and own observations made by the software management team.
\end{enumerate}

Additionally, the interviews highlighted areas for potential improvements in teams.

The first two steps (defining the software maturity model and preparing the questionnaire) have been accomplished within a few iterations: the model, its underlying conceptual framework, as well as the questionnaire, were iteratively developed in cooperation with several software teams.
The other two steps (3-4) were accomplished sequentially.

\subsection{Defining the maturity model and questionnaire}
\label{ssec:gsmm}

Maturity models are widely applied managerial instruments used for the evaluation and improvement of organisational practices and processes. The GSMM focuses on software development processes within G\'{E}ANT, considering the particular constraints in which the software teams operate. As a result, the GSMM identifies 29 Specific Goals (SGs) grouped into five Target Areas (TAs), which are essential for effective software development in G\'{E}ANT. The goals indicate objectives that need to be addressed by the software teams in the technical, organisational and human domain. 

\begin{figure}[h]
 \centering
 \includegraphics[scale=0.65]{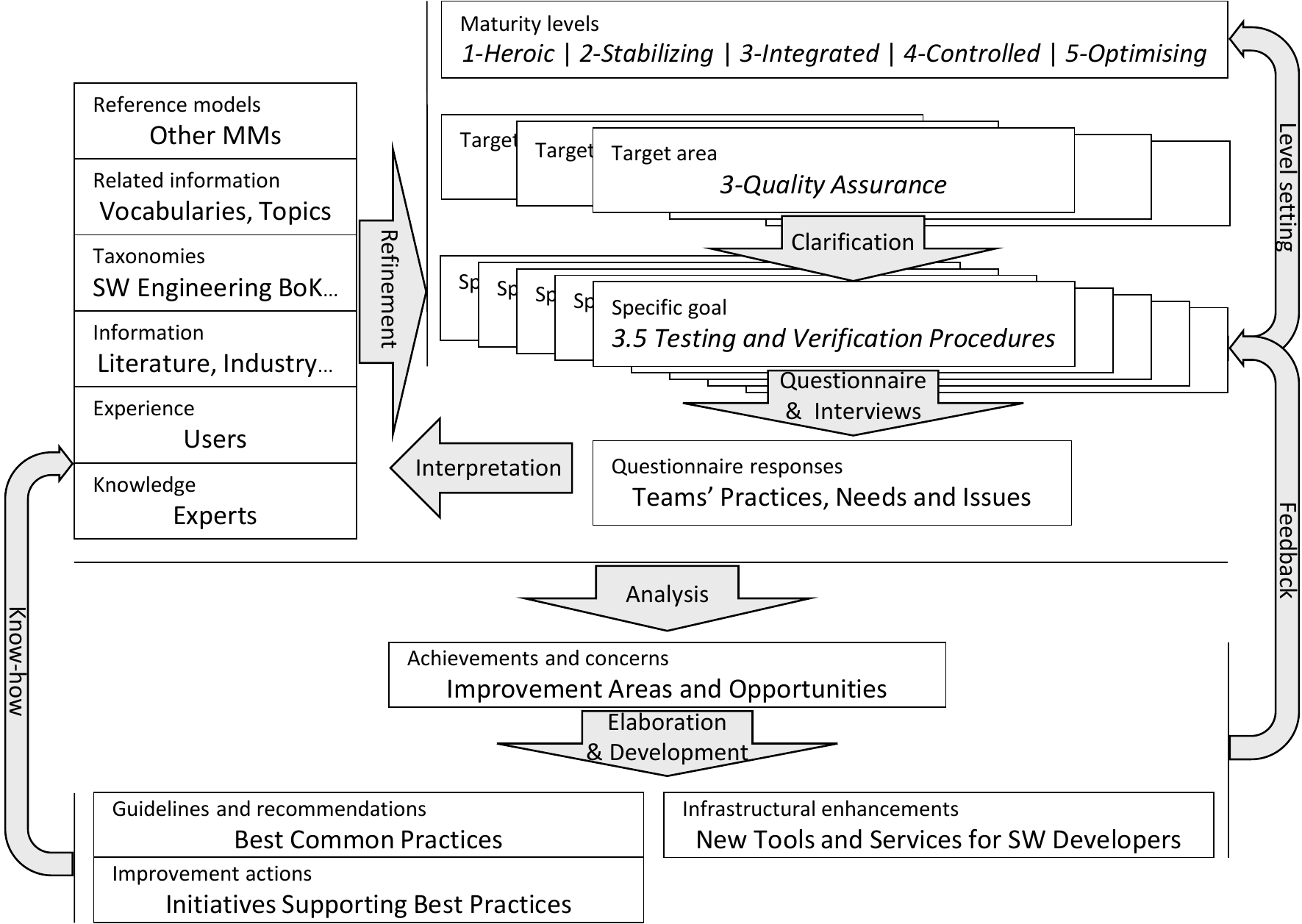}
 \caption{The overview of G\'{E}ANT Software Maturity Model development}
 \label{fig:proc}
\end{figure}

The process of defining and implementing the GSMM, its elements, produced outputs and related actions and enhancements are presented in Fig.~\ref{fig:proc}. The process of identification of specific elements of the GSMM and their further refinements is conducted iteratively, based on external and internal sources of the domain knowledge. In particular, a number of pilot interviews with the teams helped to identify key areas that are relevant in G\'{E}ANT, and further to decompose them into individual objectives. What is important, the identification process was not limited to the activities directly related to software development. We were also interested in capturing organisational, communication and human perspectives that are relevant for software teams.

This model has been elaborated in close cooperation with selected software teams, by applying iterative refinements, improvements and collecting feedback. Specifically, the process included the following phases:
\begin{enumerate}
    \item \textbf{Defining a preliminary version of GSMM}. Based on the observation of software teams, pilot interviews with the teams and a literature review, an initial version of GSMM was drafted.
    \item \textbf{Validating the preliminary version}. The preliminary version of GSMM was presented to representatives of selected teams and the executive managers responsible for supporting software development in G\'{E}ANT, to collect early feedback that could be immediately addressed.
    \item \textbf{Revising the GSMM}. Based on the collected feedback, the core elements of GSMM have been revised to better align it with the needs and practice of the teams.
    \item \textbf{Formalising the GSMM}. In parallel to that, the GSMM has been formalised into a framework by defining its elements. As a result, it can also be adapted to other application domains, beyond software development.
\end{enumerate}

\subsection{Data collection for the Common Best Practices}
\label{ssec:cbp}
To identify and capture the practices applied by the SWD teams, as well as the teams’ expectations concerning maturity, we created a questionnaire for GSMM and used as a blueprint. The questionnaire covers us meant to extract comments from the teams on each GSMM-specific goal by asking a number of closed, open and rating questions. The questions were formulated to elicit useful explanations on whether some aspects are covered or significant, how the evaluated team addresses each specific goal, what resources and approaches are used, what outputs are produced, and in which areas support may be needed. In total, we interviewed 12 teams, a representative subset of the active software teams with varied sizes.

This resulted in an extensive but targeted and streamlined process of data collection, in which the participants were able to effectively and purposefully capture their teams’ software development and management practices, but during which it was also possible to capture the teams' attitude and perception of the specific topics. 



%% file: bestpractices.tex
\section{Common Best Practices}
\label{sec:practices}

The GSMM, presented in Sec.~\ref{sec:process}, identifies the core areas and goals that contribute to the effective software development in G\'{E}ANT. However, software teams need not only the targets but also the guidance on how to organise the efforts towards meeting these them, considering the specific context of the organisation. To address this, we created a catalogue of Common Best Practices (CBPs). They have been identified based on three sources of information:
\begin{itemize}
    \item practices currently applied by the software development teams, extracted during the questionnaire interviews (see Sec.~\ref{sec:process}),
    \item recommendations provided in the literature and case studies concerning similar process optimisation efforts,
    \item direct or indirect observations made by the software management team.
\end{itemize}

The catalogue includes recommendations showing how the particular GSMM specific goals could be addressed, considering the constraints and opportunities present in G\'{E}ANT. Therefore, the best practices balance between the need for providing detailed operational guidance and giving a general direction. Both of these are extremes that would result in missing the objective of providing the teams with effective guidance that they could adapt and implement in their own practice.

\subsection{Template of a Best Practice}
\label{ssec:template}

Description of a best practice includes a large volume of diversified information of various nature and format, which can hinder its readability. To make the practices more accessible for the team members and, consequently, facilitate the adoption of practices, we decided to define a template which presents the data in a structured way. Each practice is presented as a set of attributes that describe key elements of the practice. The template comprises the following attributes:
\begin{itemize}
    \item \textbf{Objective} that the practice is expected to address;
    \item \textbf{Applicability}, describing types of projects or their phases, in which the practice could be effectively applied;
    \item \textbf{Context} that captures a specific setting (a set of technical, managerial or organisational constraints), for which the practice was identified and for which it is recommended; The practice can be used in other settings, but it may require additional validation;
    \item \textbf{Addressed elements in GSMM}, linking the practice with the SGs in the GSMM;
    \item \textbf{Prerequisites}, listing the condition necessary to apply the specific practice;
    \item \textbf{Recommendation}, outlining actions that should be undertaken to meet the SG; It includes high-level directional advice on \emph{what} to do, with lower-level details on \emph{how} to reach the goal. 
    \item \textbf{Risks}, describing possible risk factors and their consequences in case of misusing the practice;
    \item \textbf{Related practices}, indicating other practices that could be applied in a similar context;
    \item \textbf{Origin}, providing details on how the practice originated.
\end{itemize}

\subsection{Example}
\label{ssec:example}

As an example, we present a practice related to managing stakeholders that belongs to the Requirements Engineering target area. It deals with the problem of identification of relevant organisations, teams and individuals, who would affect the project or are interested in its outcomes, and properly addressing their needs and expectations.
\begin{itemize}
    \item \textbf{Objective}: Identify relevant stakeholders that can contribute to the project or have an impact on it
    \item \textbf{Context}: The practice applies to all projects.
    \item \textbf{Addressed elements in GSMM}: RE-1. Identification and overall management of stakeholders
    \item \textbf{Prerequisites}: none
    \item \textbf{Recommendation(s)}
    \begin{enumerate}
        \item Identify an initial group of stakeholders
        \begin{enumerate}
            \item Consider teams, NRENs or individuals that could be affected or could impact the project.
            \item Look for similarities to other projects, either previous or current.
            \item Look for a dominant stakeholder, who is mostly interested in the outcome of the project
        \end{enumerate}
        \item Maintain (update) the group of stakeholders
        \begin{enumerate}
            \item Publish the list of stakeholders and their representatives
            \item Periodically update (involve and retire) the group of stakeholders
            \item Apply snowballing to identify new stakeholders
            \item Categorise the stakeholders according to their relevance for the project
        \end{enumerate}
    \end{enumerate}
    \item \textbf{Risks}:
        \begin{enumerate}
            \item The identified group does not include all relevant stakeholders
            \begin{enumerate}
                \item The project may be subject to tensions, sudden changes or drifting.
                \item The decisions could be made/affected by people not officially involved in the project.
                \item The project would be not driven by stakeholders, but rather by the project team.
            \end{enumerate}
            \item Group of stakeholders is not properly updated
            \begin{enumerate}
                \item The group may not reflect the actual balance of interests.
            \end{enumerate}
        \end{enumerate}
    \item \textbf{Related practices}: BP-A-2: Create a strategy to communicate with stakeholders.  
    \item \textbf{Origin}: This practice has been defined based on the survey, supported by the observation by the software management team.
\end{itemize}
The recommendations do not provide direct instructions for the teams on how to proceed in a step-by-step manner, but rather give directional guidance. It includes advice concerning the factors and issues that facilitate addressing the respective goal in the GSMM and that should be considered by the team. As a result, the recommendations presented in a practice are a trade-off between the desire to deliver actionable procedures on one hand and the necessary abstractness on the other. The software teams are expected to analyse and adapt the recommendations to their local context.

The risk factors capture possible consequences of applying the practice improperly. In this case, they are mostly related to issues of stakeholders identification and management, including prioritisation and identification of relationships among them.

This specific practice is closely related to the process of defining the communication policy and maintaining contact with stakeholders. These two practices should be used together to ensure the maximum benefit from their implementation.

The entire catalogue is available for the G\'{E}ANT staff and currently includes 24 practices divided into five target areas that directly correspond to areas in the GSMM. Each practice addresses one or more SGs defined in the GSMM and, currently, all the goals are covered.





%% file: results.tex
\section{How to identify key areas for improvements?}
\label{sec:results}


One of the goals for analysis was to identify the areas that deserve most attention and effort. The data collected so helped in the selection of topics that should be promoted, worked on and supported through the improvement incentives. Below, we provide comments on them:
 
\begin{itemize}

    \item First, we focused on the areas in which the declared knowledge and satisfaction of the software teams was the most diversified. For those items, we can rely on the immediately available internal expertise of some teams, which reduces the effort needed for process improvement. Such harmonisation of processes among teams could additionally strengthen their collaboration.

    \item Another issue refers to areas 
    for which the survey scores were generally low or mediocre. These can be interpreted two-fold: either (1) the survey has identified a relevant topic that has been underestimated and has not been covered adequately, or (2) the topic is considered irrelevant by developers. To determine the actual status of these topics, teams and their leaders should be consulted. For that reason, it is useful at this point to establish a regular mechanism for collecting feedback from software teams. 
    The apparent difficulty associated with these topics is the insufficient internal experience, so it may be necessary to look for external expertise and support to achieve the expected improvements.

    \item The third group of interest includes the areas with uniformly high marks. Here, we can expect relatively small improvements, even if the collected data is not completely accurate. High grades given by all teams may indicate that the reached level cannot or should not be further improved.


\end{itemize}




The instruments established to collect feedback from software teams should not be used just to address the dilemmas related to the selection of improvement areas. They could be also applied to get a response about the ongoing improvement initiatives and track the metrics that are relevant for the implementation of the GSMM. However, conducting extensive interview-based surveys, like the one described in Section~\ref{sec:process}, requires significant effort; at next stages, they could be supported with simpler and frequently run online questionnaires that focus on the key elements. 

Optionally, this approach could be taken even further by associating the indicators with maturity levels. Maturity is assessed against key capability and practice characteristics linked with the selected target areas at each level of the model. Maturity levels are typically established using a five-point Likert scale where the higher the level, the higher is the level of maturity. These progressive levels guide the planning and development of roadmaps. 
Currently, there is no need for G\'{E}ANT to develop such a far-reaching maturity model. A future move in this direction would require reaching beyond G\'{E}ANT into an even larger base of software teams to capture the indicators and link their values to maturity levels.
Given the number of potential indicators and certain subjectivity in associating them with maturity levels, it is necessary to draw from the opinions and attitudes of a larger community to get non-disputable criteria. 
Although other related standardisation processes and schemes and maturity models could additionally support this development, it is crucial to establish and employ a simple, inclusive and quality process in which the indicators and levels are produced and challenged. Again, a series of simple and low-effort online surveys could be used to iteratively establish and maintain the set of used indicators and maturity criteria.


%% file: conclusions.tex
\section{Summary and conclusions}
\label{sec:conc}


The presented approach aims to establish an adaptive and practical framework for improving the quality and maturity of practices within G\'{E}ANT. It is primarily aimed at improving coordination of software process improvement efforts, fostering the collaboration of software teams, and supporting the entire software development life cycle with best practices.


In particular, the catalogue of best practices does not only help the teams to address and implement the  objectives set in GSMM but also allows them to adapt the recommended activities to their contextual constraints. As such, it fits well into the SPI Manifesto\footnote{\url{https://2019.eurospi.net/index.php/manifesto}} that emphasizes the need for embedding the maturity improvement efforts in practice and the actual needs of software teams. We believe that the catalogue of best practices will become a live, ever-growing toolbox of recommendations that would provide substantial guidance for the teams, but also receive updates from them.